\documentclass[%
aip,
rsi,%
amsmath,amssymb,
reprint,%
]{revtex4-1}
\usepackage{mathptmx}
\usepackage{pgfplots}
\pgfplotsset{compat=newest}
\usepackage{asymptote}
\usepackage{tikz}  
\usepackage{tikz-3dplot} 
\usepackage{amssymb}
\usepackage{xifthen}
\usepackage{graphicx}
\usepackage{dcolumn}
\usepackage{bm}
\pgfplotsset{every axis/.append style={
                    label style={font=\footnotesize},
                    tick label style={font=\footnotesize}  
                    }}

\begin{document}
\preprint{AIP/123-QED}
\title{Phase Field Modeling of Domain Dynamics and Polarization Accumulation in Ferroelectric HZO}

\author{Atanu K Saha}
\affiliation{School of Electrical and Computer Engineering, Purdue University, West Lafayette, IN, USA}
\author{Kai Ni}
\affiliation{Department of Electrical Engineering, University of Notre Dame, Notre Dame, IN, USA}
\author{Sourav Dutta}
\affiliation{Department of Electrical Engineering, University of Notre Dame, Notre Dame, IN, USA}
\author{Suman Datta}
\affiliation{Department of Electrical Engineering, University of Notre Dame, Notre Dame, IN, USA}
\author{Sumeet Gupta}
\affiliation{School of Electrical and Computer Engineering, Purdue University, West Lafayette, IN, USA}

\begin{abstract}
In this work, we investigate the accumulative polarization ($P$) switching characteristics of ferroelectric (FE) thin films under the influence of sequential electric-field pulses. By developing a dynamic phase-field simulation framework based on time-dependent Landau-Ginzburg model, we analyze $P$ excitation and relaxation characteristics in FE. In particular, we show that the domain-wall instability can cause different spontaneous $P$-excitation/relaxation behaviors that, in turn, can influence $P$-switching dynamics for different pulse sequences. By assuming a local and global distribution of coercive field among the grains of an FE sample, we model the $P$-accumulation process in Hf\textsubscript{0.4}Zr\textsubscript{0.6}O\textsubscript{2} (HZO) and its dependency on applied electric field and excitation/relaxation time. According to our analysis, domain-wall motion along with its instability under certain conditions plays a pivotal role in accumulative $P$-switching and the corresponding excitation and relaxation characteristics.     
\end{abstract}

\maketitle
Ferroelectric (FE) materials, particularly Zr doped HfO\textsubscript{2} (Hf\textsubscript{1-x}Zr\textsubscript{x}O\textsubscript{2}:HZO\cite{mulleracsnano2012}) have drawn significant research interest in recent times due to CMOS process compatibility\cite{pankajvlsi2017}, thickness scalability\cite{weinature2018,kimapl2018} as well as many promising attributes of ferroelectric field effect transistors\cite{sayeefacsnano2008,pankajvlsi2017} (FEFETs) for low-power logic\cite{sayeefacsnano2008,asifnature2014,asifapl2014,sahaiedm2017} and non-volatile memories\cite{trentiedm2016,chatedl2017} applications. In addition, FEFETs can provide multiple non-volatile resistive states that harness the multi-domain FE characteristics, leading to the possibilities for multi-bit synapses\cite{mulavlsitech2017,jerryiedm2017} in a neuromorphic hardware\cite{indipieee2015}. Further, newly reported accumulative polarization ($P$)-switching process\cite{mulaacsnano2018} in ultra-thin FE leads to many appealing opportunities for novel applications like correlation detection\cite{kaiiedm12018} and other non-Boolean computing paradigms\cite{mulananoscale2018}. For such emerging applications of FEFETs, the $P$-switching dynamics in response to sub/super-coercive voltage pulse trains play an important role and are, therefore, critical to understand.

To that effect, this letter analyzes spatially local $P$-switching dynamics and its participation in globally observable $P$-accumulation characteristics in response to a pulse train. Our analysis is based on a dynamic phase field model\cite{liapl2001,karin} coupled with measured accumulation characteristics of HZO. By providing the spatial distribution of $P$ ($P$-map) in different electric field (E-field) excitation and relaxation steps, we discuss different types of $P$ excitation and relaxation processes and their corresponding dependency on E-field ($E$) amplitude ($E_{max}^{app}$), ON time (or excitation time $T_{on}$) and OFF time of the pulse (or relaxation time $T_{off}$).  Finally, considering a coercive-field distribution among different FE grains, we analyze the experimental $P$-accumulation characteristics by utilizing our simulation framework.

Let us start by describing the experimentally observed $P$-switching characteristics in HZO. Fig. 1(a) shows the measured charge vs. E-field ($Q$-$E$) characteristics of a 10nm HZO film (x=0.6, grown by ALD with TiN capping layer as top and bottom contact). Here, $Q$=$P$+$\epsilon_0E$, where $\epsilon_0$ is vacuum permittivity. We observe excellent accumulative $P$-switching in HZO as the response of successive E-field stimulation (Fig. 1(b)), where the $P$-accumulation ($P^{acc}$) characteristics exhibit a strong dependence on the E-pulse properties. For example, we observe faster $P^{acc}$ with the increase in $E_{max}^{app}$ (Fig. 1(c)), increase in $T_{on}$ (Fig. 1(d)) and/or decrease in $T_{off}$ (Fig. 1(e)). Also, $P^{acc}$ saturates after a certain number of pulses. Such saturation occurs at higher $P$ with the increase in $E_{max}^{app}$, increase in $T_{on}$ and decrease in $T_{off}$. It is noteworthy that such accumulated-$P$ observed in the experiments is the average of locally accumulated-$P$ in different grains. Hence, to explain the experimental results described above, it is critical to understand the spatially local $P$-switching dynamics in an individual grain. We analyze such processes in detail based on our phase-field model, calibrated to the experiment.
\begin{figure}
    \centering
    \includegraphics[scale=0.895]{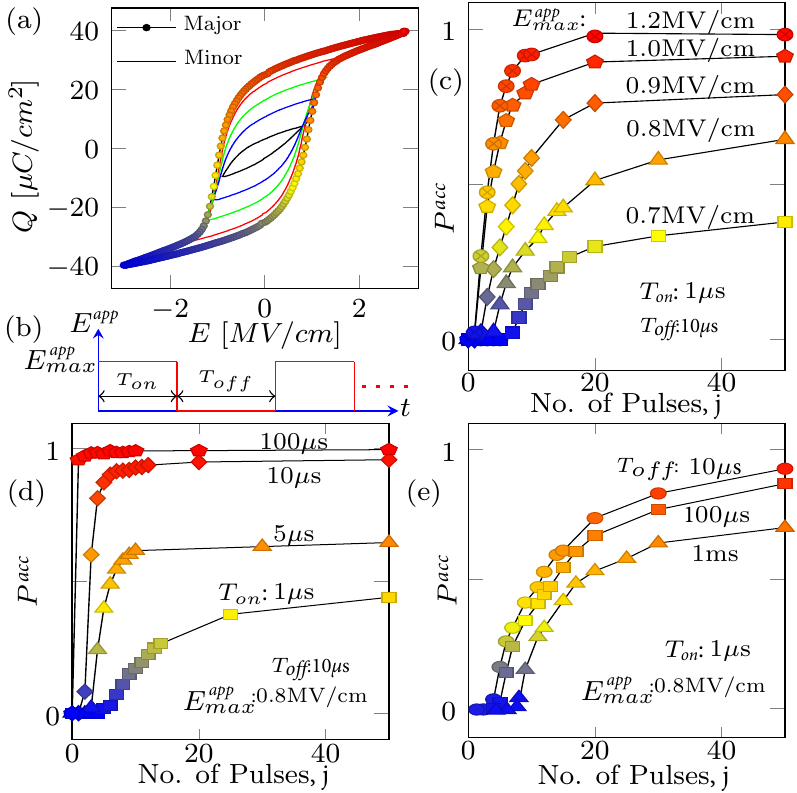}
    \caption{\footnotesize(a) $Q$-$E$ curves of a 10nm HZO film. (b) Applied E-field pulses showing pulse-amplitude ($E_{max}^{app}$), excitation time ($T_{on}$) and relaxation time ($T_{off}$). Accumulated polarization, $P^{acc}$ (=$(P+P_r)/2P_r$, $P_r=25\mu C/cm^2$) vs. number of E-field pulses ($j$) for different (c) $E_{max}^{app}$, (d) $T_{on}$ and (e) $T_{off}$.}
    \label{fig:my_label}
\end{figure}

Considering a thin ($\sim$10nm) FE film, we assume that the $P$ direction is only along the thickness ($z$-axis). Hence, $P_x$=0, $P_y$=0, $P_z\neq$0 and $P_z$ can have a spatial distribution in $x$-$y$ plane. However, we assume uniform $P_z$ along the $z$-axis ($dP_z/dz$=0) owing to the ultra-thin nature of the film. The temporal and spatial evolution of $P$ can be described by the time ($t$)-dependent Landau-Ginzburg (TDLG) equation\cite{karin,Nambu}:
${\delta F}/{\delta P_z} = -\rho({\partial P_z}/{\partial t})$. Here, $\rho$ is the kinetic coefficient and $F$ is the total energy\cite{Nambu} of the system composed of free energy, domain-wall energy, electrostatic and electroelastic energy. Considering up to the 6\textsuperscript{th} order terms in Landau's free energy expansion\cite{bell}, the normalized representation of TDLG equation within the FE is given by the following equation.
\begin{eqnarray}
-\rho_n \frac{\partial P_n}{\partial t} = -K_P^n\nabla^2P_n-E_n^{app}+\hat{\alpha} P_n + \hat{\beta} P_n^3 + \hat{\gamma} P_n^5 ~
\label{eq: main}
\end{eqnarray}
Here, $P_n (=P_z/P_C$) and $E_n^{app} (=E^{app}/E_C$) are polarization and E-field normalized with respect to $E_c$ (coercive field) and $P_C$ ($P$ at $E$=$E_C$), respectively.  $\hat{\alpha}$, $\hat{\beta}$ and $\hat{\gamma}$ are the normalized Landau coefficients, $\rho_n$ is the normalized kinetic coefficient and $K_{P}^n$ is the normalized domain-interaction parameter. Eqn. 1, is similar to the Euler-Lagrange equation of motion\cite{ken}, incorporates interactions among neighboring domains (via $K_{P}^n\nabla^2P_n$) giving rise to Klein-Gordon type field equation\cite{Band}. In our simulations, we self-consistently solve eqn. 2 in a real space grid by considering Neumann boundary at the edges\cite{cano}. We include a comprehensive discussion on parameter extraction, normalization and simulation methodology in the supplementary section. It is noteworthy that $P_n$ denotes normalized microscopic $P$ in each grid point, while the analogous quantity of experimentally measured $P$ is spatial average of $P_n$, denoted as $\bar{P}_n$. Also, $K_P^n\nabla^2P_n$ can be thought of as the local effective interaction E-field, $E_n^{int}$. Therefore, the $P$ switching depends on $E_n^{app}$+$E_n^{int}$. For instance, $P$-switching will occur for $E_n^{app}$+$E_n^{int}>$1 (since the normalized coercive field=1). 

\begin{figure}
    \centering
   \includegraphics[scale=0.78]{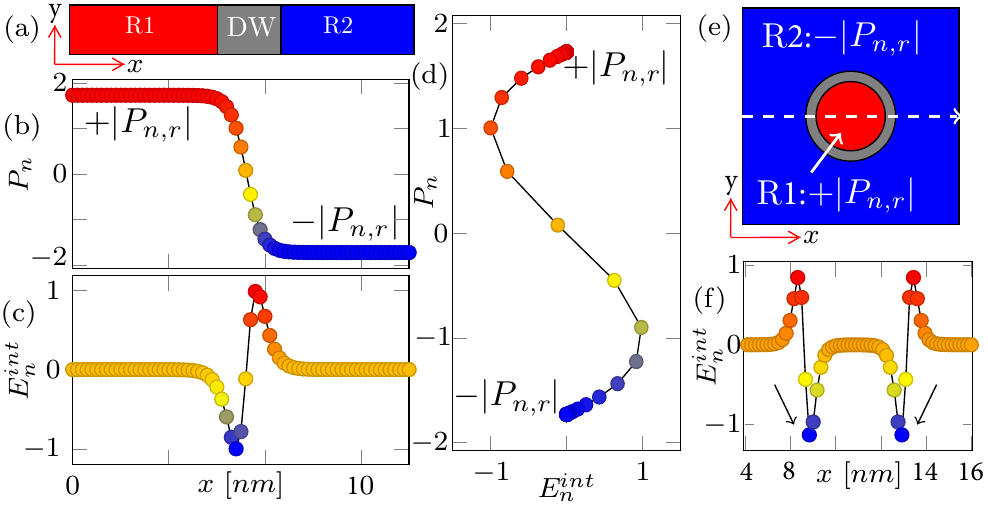}
    \caption{\footnotesize(a) FE structure showing 1D domain-wall (DW) and spatial distribution of (b) polarization, $P_n$ and (c) interaction E-field, $E_n^{int}$. (d) Static $P_n$-$E_n^{int}$ relation. (e) FE structure showing 2D DW and spatial distribution of (f) $E_n^{int}$ along the $x$-axis.}
    \label{fig:my_label}
\end{figure}
\begin{figure}
    \centering
    \includegraphics[scale=0.86]{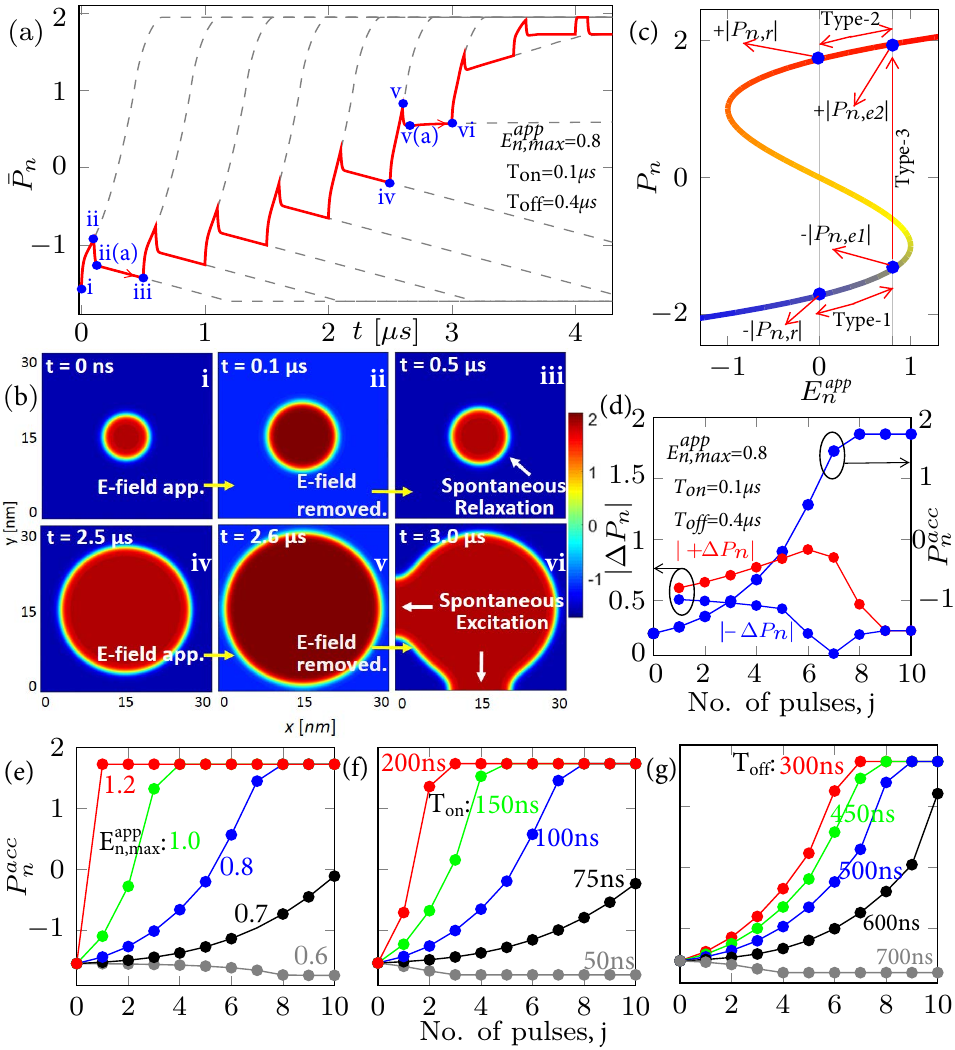}
    \caption{\footnotesize(a) Simulated transient $\bar{P}_n$ considering case-1 for a sequence of E-field pulses. Corresponding (b) $P$-map at point i-vi. (c) Static $P_n$ vs. $E_{n}^{app}$ showing different stimulated excitation/relaxation components. (d) Increase and decrease in $P_n$ (respectively $|\Delta^+P_n|$ and $|\Delta^-P_n|$) and accumulated $P$ ($P_n^{acc}$) in each excitation/relaxation sequence w.r.t. pulse number ($j$).  $P_n^{acc}$ vs. $j$ for different (e) $E_{n,max}^{app}$, (f) $T_{on}$ and (g) $T_{off}$.} 
    \label{fig:my_label}
\end{figure}
In general, $P$-switching can take place in two different ways, namely (i) direct nucleation and (ii) Domain-wall (DW) assisted nucleation. To understand such $P$-switching processes, let us start by considering the FE sample in Fig. 2(a), where region R1 exhibits $P_n=+|P_{n,r}|$ and R2 exhibits $P_n=-|P_{n,r}|$ and they are separated by a DW within which $P_n$ varies gradually along the $x$-axis (Fig. 2(b)). Here, $P_{n,r}$ is the remnant polarization. In this case, the domain structure is effectively 1D as $d^2P_n/dy^2=0$. Direct nucleation occurs for super-coercive applied fields ($|E_n^{app}|>1$), wherein region R2 will switch to $+P$ at once if E-field is applied for a sufficient time. On the other hand, DW assisted nucleation (which is the main focus of this work) is observed for sub-coercive applied fields ($|E_n^{app}|<1$), in which $E_n^{int}$ plays a key role. To explain this, let us consider $E_n^{app}=0$ and static condition ($dP_n/dt$=0). Hence, eqn. (1) can be written as, $E_n^{int}$=$K_P^n\nabla^2P_n$=$\hat{\alpha}_n P_n$+$\hat{\beta}_n P_n^3$+$\hat{\gamma}_n P_n^5$. Note, $E_n^{int}$ is localized and non-zero only within DW (Fig. 2(c)). Fig. 2(d) shows the relation between $E_n^{int}$ and $P_n$, signifying that the symmetric spatial distribution of $P_n$ provides a symmetric $E_n^{int}$ in the 1D case. Here, the symmetric $E_n^{int}$ plays a critical role by balancing the forces due to $E_n^{int}$ on the two sides of the DW, yielding stable and static DW for $E_n^{app}$=0. This can also be understood by noting that $E_n^{int}\leq$1, which leads to stable DW due to no $P$-switching. However, by applying a sub-coercive E-field, $0<|E_n^{app}|<1$, we can get a total local E-field, $|E_n^{app}$+$E_n^{int}|>$1 and that can eventually initiate $P$-switching. Such $P$-switching is a spatially local and temporally gradual process, which is referred to DW motion or DW assisted nucleation. (For more details on 1D $P$-switching, see supplementary section).

With the understanding of the 1D $P$-switching, let us now consider 2D FE dynamics for which, we analyze two cases by considering homogeneous $E_C$ (case-1) and a distribution of $E_C$ (case-2) in an FE grain. Let us start with case-1 and consider a square FE sample (Fig. 2(e)) where a circular region R1 exhibits $P_n=+|P_{n,r}|$, which is surrounded by R2 with  $P_n=-|P_{n,r}|$. Here, the DW is 2D and the corresponding $\nabla^2P_n$ in the polar coordinate can be written as $[(\partial^2P_n/\partial r^2)+(1/r)(\partial P_n/\partial r)]$. Note that, the $(1/r)(\partial P/\partial r)$ term exhibits a radial dependency and hence, $\nabla^2P_n$ becomes radially asymmetric, even for a symmetric radial distribution of $P_n$ (symmetric $\partial^2P_n/\partial r^2$ and $\partial P_n/\partial r$). Therefore, $E_n^{int}$ becomes spatially asymmetric (Fig. 3(f)), where $|E_n^{int}|>1$ at the inner interface and $|E_n^{int}|<1$ at the outer interface of DW. Such asymmetry in the $E_n^{int}$ causes the DW to undergo an effective inward force. Consequently, the DW becomes unstable and R1 region shrinks spontaneously with time (as $|E_n^{int}|>1$ at the interior end of the DW). Such spontaneous phenomena play an important role in the $P$-switching dynamics that we discuss subsequently. (Note, such a DW instability is in contrast with the 1D case that we discussed above where symmetric $E_n^{int}$ leads to stable DW).

Let us now consider a sequence of sub-coercive E-field pulses ($E_{n,max}^{app}$=0.8) applied to this sample of FE. Simulated transient $\bar{P}_n$ is shown in Fig. 3(a) and the initial $P$-map at $t$=0ns is shown in Fig. 3(b)-i, where the initially switched region (red) can be assumed as a pinned domain. After the arrival of first E-field pulse, R1 domain grows circularly, nucleating new lattices sequentially at the outer edge of the DW (Fig. 3(b)-ii). That implies an increase in R1 area and decrease in R2 area by an amount $\Delta A_j^+$ ($j$= E-field pulse number). The corresponding $P$-excitation characteristics (Fig. 3(a):(i-ii)) comprise of three different components (Fig. 3(c)), \textit{i.e.} type-1: $-|P_{n,r}|$ to $-|P_{n,e1}|$ (in R2), type-2: $+|P_{n,r}|$ to $+|P_{n,e2}|$ (in R1) and type-3: $-|P_{n,r}|$ to $+|P_{n,e2}|$ (in $\Delta A_j^+$). 

After the end of first E-field pulse, the DW propagation stops and the $P$ changes due to type 1 and 2 components get immediately relaxed to $-|P_{n,r}|$ and $+|P_{n,r}|$, respectively. We call these type-1 and 2 relaxations, respectively (Fig. 3(c)). Similarly, the newly nucleated part ($\Delta A_j^+$) also rapidly get relaxed to $+|P_{n,r}|$ by following type-2 relaxation. Corresponding transient relaxation in $\bar{P}_n$ can be seen in Fig. 3(a) (from point ii to ii(a)). Interestingly, followed by such rapid relaxation, there is another relaxation component that gradually reduces $\bar{P}_n$ until the arrival of next E-field pulse (Fig. 3(a): ii(a)-iii). Such spontaneous $P$-relaxation is the outcome of DW instability (due to $E_n^{int}$ asymmetry – discussed above) that causes spontaneous shrinking of R1 domain (Fig. 3(b): ii-iii). Let us define the spontaneous decrease in R1 area in the absence of E-field as $\Delta A_j^-$.   

Now, due to sequential E-field pulses, rather than completely collapsing, DW moves further towards the grain boundary by following $P$-excitation and relaxation sequences. Once DW reaches sufficiently close to the grain boundary, spontaneous relaxation is not observed in the absence of E-field (during $T_{off}$). Instead, spontaneous $P$-excitation (Fig. 3(a):v(a)-vi) takes place. This is because, as the DW reaches near the edges, R2 domain becomes very narrow and hence, $\nabla^2P_n$ increases. Therefore, at the outer interface of the DW (alongside R2), $|E_n^{int}|>1$ and that causes an effective outward force in R2. As a result, R2 domain spontaneously switches to $+P$ (Fig. 3(b): iv-vi). After all the lattices switch to $+|P_{n,r}|$, transient $\bar{P}_n$ exhibits only type-2 excitation and relaxation.

\begin{figure}
    \centering
    \includegraphics[scale=0.86]{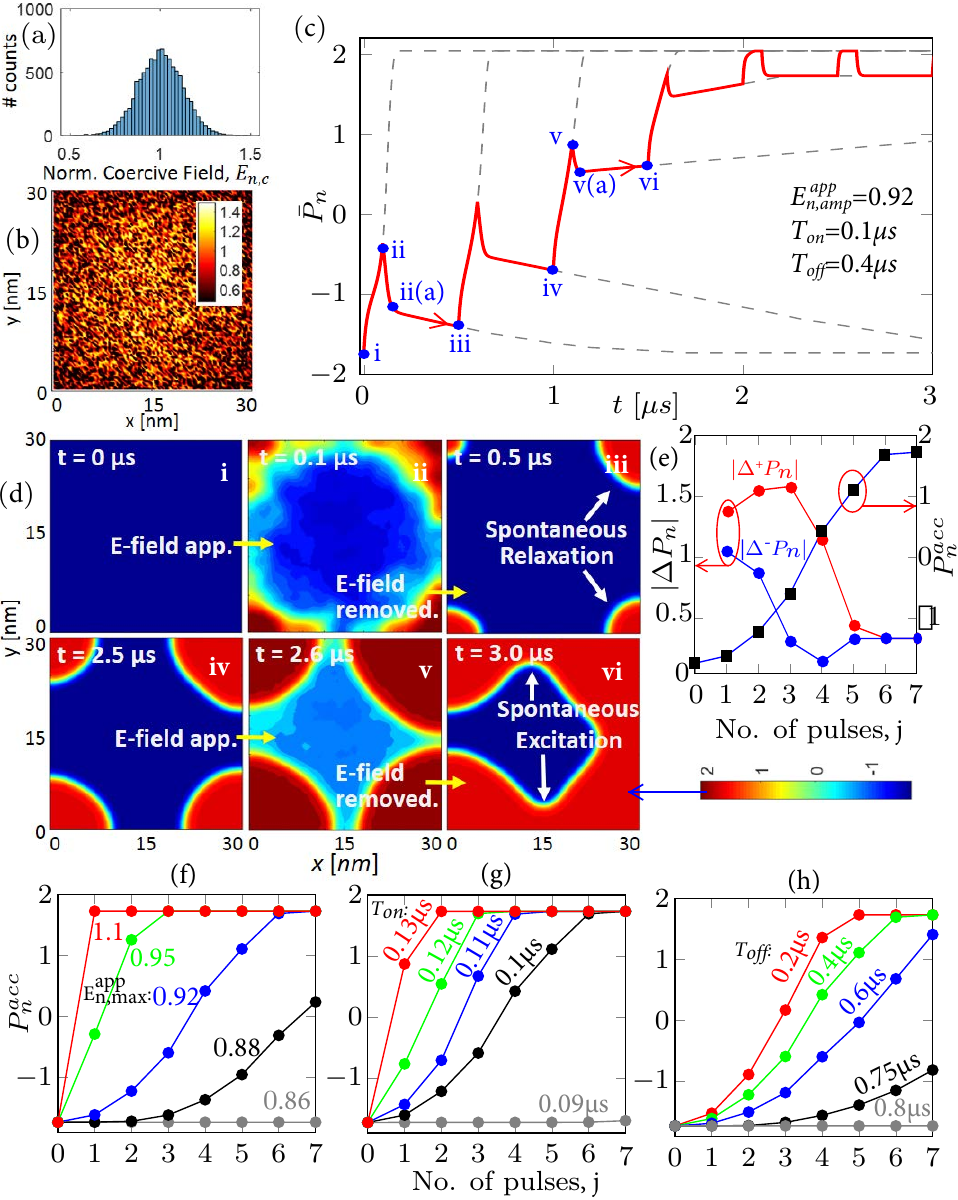}
    \caption{\footnotesize Distribution of $E_{n,C}$, (a) as area fraction and (b) as spatial, in an FE grain. (c) Simulated transient $\bar{P}_n$. Corresponding (d) $P$-map at point i-vi, (e) $|\Delta^+P_n|$, $|\Delta^-P_n|$ and $P_n^{acc}$ w.r.t. $j$. $P_n^{acc}$ vs. $j$ for different (f) $E_{n,max}^{app}$, (g) $T_{on}$ and (h) $T_{off}$.}
    \label{fig:my_label}
\end{figure}
Note that the $P$-maps at the beginning of different E-field pulses are different and play an important role in each $P$-excitation/relaxation behavior. Let us define the increase in $\bar{P}_n$ during each excitation period ($|\Delta^+ \bar{P}_n|_j$). Recall that each $|\Delta^+ \bar{P}_n|_j$ consists of three excitation components. It can be shown mathematically from eqn. 1 that $P$ change due to type-1 excitation is higher in magnitude than type-2 (Fig. 3(c)). Now, with the increase in pulse number ($j$), R1 area increases and R2 area decreases. Therefore, the contribution from type-2 excitation (in R1) increases and type-1 excitation (in R2) decreases. Hence, we expect an overall decrease in total excitation (type-1+2) with respect to (w.r.t.) $j$. Now, type-3 excitation  (corresponding to R1 area increase during $j$-th excitation) can be written as, $\Delta A_j^+$=$\pi[(r_j+dr_j)^2-r_j^2]$=$\pi[dr_j^2+2r_jdr_j]$. Here $r_j$ is the domain radius before $j$-th excitation and $dr_j$ is the increase in radius during $j$-th excitation. Note that a linear increase in $dr_j$ w.r.t. time gives rise to quadratic increase in R1 area. That implies that if we keep the E-pulse ON for a long time, the $\bar{P}_n$ dynamics will be quadratic w.r.t. time (gray lines in Fig. 3(a)). Similarly, assuming $dr_j$ as constant irrespective of the value of $j$, we can see that $\pi\times2r_jdr_j$ increases with $j$ as $r_j$ increases. That implies an incremental change in $\Delta A_j^+$ and hence, type-3 excitation component increases with $j$. Note that the type-3 contribution is dominant over type-1+2 and therefore, $|\Delta^+ \bar{P}_n|_j$ increase with the increase with $j$ up to $j$=6 (Fig. 3(d)). For, $j>$6, R1 domain reaches the grain boundary and the quadratic growth of R1 no longer holds true. Hence, type-3 contribution decreases significantly, leading to the domination of type-1+2 excitation and decrease in $|\Delta^+\bar{P}_n|_j$ with $j$. After R2 domain vanishes (or switched to $+P$ at $j$=9), $|\Delta^+\bar{P}_n|_j$ only exhibits type-2 excitation.     

Similarly, the decrease in $\bar{P}_n$ during each relaxation period ($|\Delta^-\bar{P}_n|_j$) consists of type-1+2 relaxation and a spontaneous part. Like type-1+2 excitation, type-1+2 relaxation decreases as $j$ increases. However, the spontaneous component ($\Delta A_j^-$) behaves non-monotonically w.r.t. $j$. $\Delta A_j^-$ (decrease in R1 area) changes sign from positive ($+$) to negative ($-$) at $j$=6 as the spontaneous component changes from relaxation to excitation characteristics. Therefore, till the spontaneous contribution is relaxation ($j$$\leq$$6$), $|\Delta^-\bar{P}_n|_j$ decreases with the increase in $j$. Once the spontaneous contribution leads to excitation ($j$$>$$7$), $|\Delta^-\bar{P}_n|_j$ increases with $j$ and becomes constant at $j$=9 (with only type-2 relaxation in R1).  

$\bar{P}_n$ at the end of each excitation-relaxation sequence, called accumulated polarization ($P_n^{acc}$) is shown in Fig. 3(a). Note that the change in $P_n^{acc}$ at each pulse is basically proportional to $\Delta A_j^+$-$\Delta A_j^-$. We discussed earlier that $\Delta A_j^+$ shows incremental increase with the increase in $j$, whereas, $\Delta A_j^-$ exhibits non-monotonic change along with a sign change from `+' to `-'. Therefore, $P_n^{acc}$ initially increases slowly when $\Delta A_j^-$ is `+' and once $\Delta A_j^-$ becomes `-', then increases rapidly. On the other hand, the flat region (Fig. 3(a), $j\geq$9) in $P_n^{acc}$ signifies an absence in $P$-accumulation once the whole sample (or grain) completely switches to $+P$. 

The trends in $P$ accumulation w.r.t the pulse attributes are illustrated in Fig. 3(e-g).  With the increase in pulse amplitude ($E_{n,max}^{app}$), R1 domain grows more rapidly (increase in $dA^+/dt$) leading to faster accumulation (Fig. 3(e)). Similarly, with the increase in $T_{on}$, $\Delta A_j^+$ increases during each pulse and therefore, $P_n^{acc}$ saturates at a lower $j$ (Fig. 3(f)). Also, an increase in $T_{off}$ leads to an increase in spontaneous relaxation (increase in $\Delta A_j^-$). Consequently, larger number of pulses is required for $P_n^{acc}$ to get saturated (Fig. 3(g)). Note that, if $E_{n,max}^{app}$ or/and $T_{on}$ is/are very low, so that $(\Delta A_j^+$-$\Delta A_j^-)<0$, then, rather than accumulation, R1 can get completely relaxed to $-P$. The same is true for a high $T_{off}$. Such scenarios can be seen in Fig. 3(e-g) (gray lines). Note that here, we assume the pinned-type (or initially nucleated) domain at the center of the grain. However, such pinned domain can be at random positions within the grain. In such cases too, the trends in $P$-switching dynamics remain the same as discussed (see supplementary section).

Now, we consider case-2, where we assume a Gaussian distribution of $E_C$ in an FE grain (Fig.4(a)) by considering a spatial distribution of $\hat{\alpha}$, $\hat{\beta}$ and $\hat{\gamma}$ (see supplementary section). Note that $E_C$ is assumed to be less near the grain boundary (Fig. 4(b)), which can be understood as the cause of strain relaxation near the edges\cite{Emelyanov2002,joo_1999}. Like the previous discussion, considering a sequence of E-field pulses ($E_{n,max}^{app}$=0.92), simulated $\bar{P}_n$ is shown in Fig. 4(c). Note that the FE grain was initially switched to $-|P_{n,r}|$ (Fig. 4(d): i). Therefore, $P$-switching occurs as a two-step process: (1) E-field induced nucleation and (2) E-field assisted domain growth. Once the first E-field pulse arrives, direct nucleation starts from the grain edges (with lower $E_C$) and propagates inward (Fig. 4(d): ii). After the end of E-field pulse, further nucleation stops and type-1-2 relaxation takes place (Fig. 4(c):ii-ii(a)) followed by a spontaneous relaxation (Fig. 4(c):ii(a)-iii and Fig. 4(d): iii) due to DW instability. However, after the 3rd pulse ($j>$3), spontaneous excitation occurs in the absence of E-field (Fig. 4(c): v(a)-vi), rather than spontaneous relaxation. Note that, the origin of spontaneous excitation in this case is not the instability of DW near the grain boundary. In contrast, when two DWs are sufficiently close, then the intermediate domain experiences a non-zero $E_n^{int}$ governed by both the DWs. Therefore, total $E_n^{int}>1$ at DW interfaces alongside the intermediate domain. Consequently, the intermediate domain becomes unstable and spontaneously switches to $+P$ (Fig. 4(d): v-vi). Such spontaneous excitation continues up to $j$=4, till all the lattices have switched to $+P$. Corresponding $|\Delta^+\bar{P}_n|_j$, $|\Delta^-\bar{P}_n|_j$ and $P_n^{acc}$ are shown in Fig. 4(e-h) that present similar trends like case-1. However, an important difference between these two cases is stronger spontaneous excitation and relaxation in case-2 compared to case-1 (see supplementary section for details), which yields relatively abrupt $P$-switching in case-2. 

With the understating of $P$-excitation/relaxation processes in an FE grain, we now analyze the $P$-accumulation in HZO by considering an ensemble of grains. The global $E_C$ distribution for HZO (80$\mu$m$\times$80$\mu$m) is shown in Fig. 5(a), which we extract from the measured $P$-$E$ curves (discussed in supplementary section). Then, we use each sampled $E_C$ as the mean value of a local Gaussian distribution of $E_C$ in a grain (like case-2). Considering a large number of grains and multiplying each local $E_C$ distribution with the corresponding area fraction, the resultant global distribution of $E_C$ is shown in Fig. 5(a).

\begin{figure}
    \centering
   \includegraphics[scale=0.8]{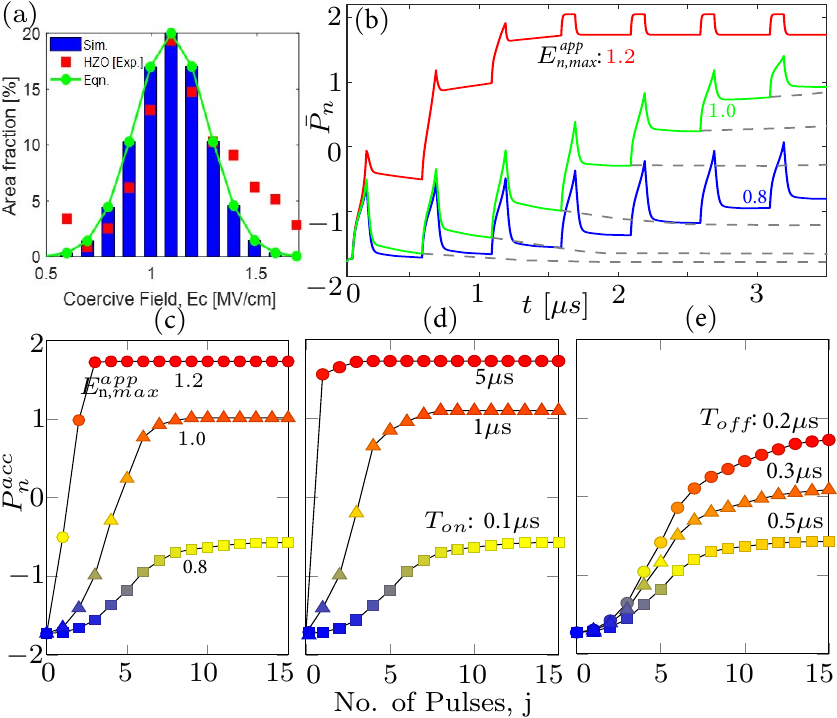}
    \caption{\footnotesize(a) $E_C$ distribution in HZO: (red) experiment; (blue/green) used for simulation. (b) Simulated transient $\bar{P}_n$ for different $E_{n,max}^{app}$. $P_n^{acc}$ vs. $j$ for different (c) $E_{n,max}^{app}$, (d) $T_{on}$ and (e) $T_{off}$.}
    \label{fig:my_label}
\end{figure}
By considering a sequence of E-field pulses (for $E_{n,max}^{app}$=0.8,1.0,1.2), simulated $\bar{P}_n$ and corresponding $P_n^{acc}$ are shown in Fig. 5(b-c). While, the signatures of the dynamics of single grain (discussed above) are manifested in HZO (ensemble of grains), two important differences can be observed in HZO: (1) saturation of accumulated $P$ occurs at an intermediate value which increases for higher $E_{n,max}^{app}$, higher $T_{on}$ and or lower $T_{off}$ (Fig. 5(c-e)) and (2) for a long relaxation time, the overall $P$ does not relax completely (Fig.5 (b): gray dashed lines).  The former observation is attributed to two processes. First, grains with low mean $E_C$ switch completely after sufficient number of pulses and therefore, do not contribute to $P$ accumulation further, leading to intermediate saturation.  Second, grains with sufficiently high mean $E_C$ exhibit  low initial nucleation for a given $E_{n,max}^{app}$ and $T_{on}$. Therefore, given a relaxation time, the grains with higher $E_C$ are more likely to relax completely and hence, do not participate in $P$-accumulation. Now, with the increase in $E_{n,max}^{app}$ and $T_{on}$, initial nucleation is enhanced, reducing the probability of complete relaxation in high $E_C$  grains . A decrease in $T_{off}$ also has a similar effect on relaxation. This results in the contribution of larger number of grains to $P$-accumulation leading to $P$ saturation at a higher value (Fig. 5(c-e)). For the second observation (incomplete relaxation for large $T_{off}$), the reason is attributed to low mean $E_C$ grains that completely switch during the excitation and hence, do not participate in spontaneous relaxation. Note that the large distribution of $E_C$ corresponds to the large area of our fabricated HZO sample. However, by scaling the area of HZO, less number of grains along with a compact global distribution of $E_C$ can be achieved. Therefore, the $P$-accumulation of a scaled HZO should exhibit less number of saturation levels as well as more abrupt $P$-switching.    

In summary, we experimentally demonstrated the accumulative $P$-switching in HZO. Then, developing a phase-field model, we discuss the $P$-switching dynamics by analyzing different stimulated and spontaneous $P$-excitation/relaxation mechanisms governed by domain-domain interaction and DW instability. We attribute the strength and directional change in DW instability as one of the key factors for accumulative $P$-switching. Finally, considering an inter/intra-grain coercive-field distribution in our simulations, we describe the experimentally observed accumulative $P$-switching in HZO and their dependence on E-field pulse attributes.

\bibliography{aipsamp}

\begin{thebibliography}{25}%
\makeatletter
\providecommand \@ifxundefined [1]{%
 \@ifx{#1\undefined}
}%
\providecommand \@ifnum [1]{%
 \ifnum #1\expandafter \@firstoftwo
 \else \expandafter \@secondoftwo
 \fi
}%
\providecommand \@ifx [1]{%
 \ifx #1\expandafter \@firstoftwo
 \else \expandafter \@secondoftwo
 \fi
}%
\providecommand \natexlab [1]{#1}%
\providecommand \enquote  [1]{``#1''}%
\providecommand \bibnamefont  [1]{#1}%
\providecommand \bibfnamefont [1]{#1}%
\providecommand \citenamefont [1]{#1}%
\providecommand \href@noop [0]{\@secondoftwo}%
\providecommand \href [0]{\begingroup \@sanitize@url \@href}%
\providecommand \@href[1]{\@@startlink{#1}\@@href}%
\providecommand \@@href[1]{\endgroup#1\@@endlink}%
\providecommand \@sanitize@url [0]{\catcode `\\12\catcode `\$12\catcode
  `\&12\catcode `\#12\catcode `\^12\catcode `\_12\catcode `\%12\relax}%
\providecommand \@@startlink[1]{}%
\providecommand \@@endlink[0]{}%
\providecommand \url  [0]{\begingroup\@sanitize@url \@url }%
\providecommand \@url [1]{\endgroup\@href {#1}{\urlprefix }}%
\providecommand \urlprefix  [0]{URL }%
\providecommand \Eprint [0]{\href }%
\providecommand \doibase [0]{http://dx.doi.org/}%
\providecommand \selectlanguage [0]{\@gobble}%
\providecommand \bibinfo  [0]{\@secondoftwo}%
\providecommand \bibfield  [0]{\@secondoftwo}%
\providecommand \translation [1]{[#1]}%
\providecommand \BibitemOpen [0]{}%
\providecommand \bibitemStop [0]{}%
\providecommand \bibitemNoStop [0]{.\EOS\space}%
\providecommand \EOS [0]{\spacefactor3000\relax}%
\providecommand \BibitemShut  [1]{\csname bibitem#1\endcsname}%
\let\auto@bib@innerbib\@empty
\bibitem [{\citenamefont {Muller}\ \emph {et~al.}(2012)\citenamefont {Muller},
  \citenamefont {Boscke}, \citenamefont {Schroder}, \citenamefont {Mueller},
  \citenamefont {Brauhaus}, \citenamefont {Bottger}, \citenamefont {Frey},\
  and\ \citenamefont {Mikolajick}}]{mulleracsnano2012}%
  \BibitemOpen
  \bibfield  {author} {\bibinfo {author} {\bibfnamefont {J.}~\bibnamefont
  {Muller}}, \bibinfo {author} {\bibfnamefont {T.~S.}\ \bibnamefont {Boscke}},
  \bibinfo {author} {\bibfnamefont {U.}~\bibnamefont {Schroder}}, \bibinfo
  {author} {\bibfnamefont {S.}~\bibnamefont {Mueller}}, \bibinfo {author}
  {\bibfnamefont {D.}~\bibnamefont {Brauhaus}}, \bibinfo {author}
  {\bibfnamefont {U.}~\bibnamefont {Bottger}}, \bibinfo {author} {\bibfnamefont
  {L.}~\bibnamefont {Frey}}, \ and\ \bibinfo {author} {\bibfnamefont
  {T.}~\bibnamefont {Mikolajick}},\ }\href {\doibase 10.1021/nl302049k}
  {\bibfield  {journal} {\bibinfo  {journal} {Nano Lett.}\ }\textbf {\bibinfo
  {volume} {12}},\ \bibinfo {pages} {4318} (\bibinfo {year}
  {2012})}\BibitemShut {NoStop}%
\bibitem [{\citenamefont {Sharma}\ \emph {et~al.}(2017)\citenamefont {Sharma},
  \citenamefont {Tapily}, \citenamefont {Saha}, \citenamefont {Zhang},
  \citenamefont {Shaughnessy}, \citenamefont {Aziz}, \citenamefont {Snider},
  \citenamefont {Gupta}, \citenamefont {Clark},\ and\ \citenamefont
  {Datta}}]{pankajvlsi2017}%
  \BibitemOpen
  \bibfield  {author} {\bibinfo {author} {\bibfnamefont {P.}~\bibnamefont
  {Sharma}}, \bibinfo {author} {\bibfnamefont {K.}~\bibnamefont {Tapily}},
  \bibinfo {author} {\bibfnamefont {A.~K.}\ \bibnamefont {Saha}}, \bibinfo
  {author} {\bibfnamefont {J.}~\bibnamefont {Zhang}}, \bibinfo {author}
  {\bibfnamefont {A.}~\bibnamefont {Shaughnessy}}, \bibinfo {author}
  {\bibfnamefont {A.}~\bibnamefont {Aziz}}, \bibinfo {author} {\bibfnamefont
  {G.~L.}\ \bibnamefont {Snider}}, \bibinfo {author} {\bibfnamefont
  {S.}~\bibnamefont {Gupta}}, \bibinfo {author} {\bibfnamefont {R.~D.}\
  \bibnamefont {Clark}}, \ and\ \bibinfo {author} {\bibfnamefont
  {S.}~\bibnamefont {Datta}},\ }in\ \href {\doibase
  10.23919/VLSIT.2017.7998160} {\emph {\bibinfo {booktitle} {2017 Symposium on
  VLSI Technology}}}\ (\bibinfo {year} {2017})\ pp.\ \bibinfo {pages}
  {T154--T155}\BibitemShut {NoStop}%
\bibitem [{\citenamefont {Wei}\ \emph {et~al.}(2014)\citenamefont {Wei},
  \citenamefont {Nukala}, \citenamefont {Salverda}, \citenamefont {Matzen},
  \citenamefont {Zhao}, \citenamefont {Momand}, \citenamefont {Everhardt},
  \citenamefont {Agnus}, \citenamefont {Blake}, \citenamefont {Lecoeur},
  \citenamefont {Kooi}, \citenamefont {Íñiguez}, \citenamefont {Dkhil},\ and\
  \citenamefont {Noheda}}]{weinature2018}%
  \BibitemOpen
  \bibfield  {author} {\bibinfo {author} {\bibfnamefont {Y.}~\bibnamefont
  {Wei}}, \bibinfo {author} {\bibfnamefont {P.}~\bibnamefont {Nukala}},
  \bibinfo {author} {\bibfnamefont {M.}~\bibnamefont {Salverda}}, \bibinfo
  {author} {\bibfnamefont {S.}~\bibnamefont {Matzen}}, \bibinfo {author}
  {\bibfnamefont {H.~J.}\ \bibnamefont {Zhao}}, \bibinfo {author}
  {\bibfnamefont {J.}~\bibnamefont {Momand}}, \bibinfo {author} {\bibfnamefont
  {A.~S.}\ \bibnamefont {Everhardt}}, \bibinfo {author} {\bibfnamefont
  {G.}~\bibnamefont {Agnus}}, \bibinfo {author} {\bibfnamefont {G.~R.}\
  \bibnamefont {Blake}}, \bibinfo {author} {\bibfnamefont {P.}~\bibnamefont
  {Lecoeur}}, \bibinfo {author} {\bibfnamefont {B.~J.}\ \bibnamefont {Kooi}},
  \bibinfo {author} {\bibfnamefont {J.}~\bibnamefont {Íñiguez}}, \bibinfo
  {author} {\bibfnamefont {B.}~\bibnamefont {Dkhil}}, \ and\ \bibinfo {author}
  {\bibfnamefont {B.}~\bibnamefont {Noheda}},\ }\href {\doibase
  10.1038/s41563-018-0196-0} {\bibfield  {journal} {\bibinfo  {journal} {Nature
  Materials}\ }\textbf {\bibinfo {volume} {17}},\ \bibinfo {pages} {1095}
  (\bibinfo {year} {2014})}\BibitemShut {NoStop}%
\bibitem [{\citenamefont {Khan}\ \emph {et~al.}(2018)\citenamefont {Khan},
  \citenamefont {Bhowmik}, \citenamefont {Yu}, \citenamefont {Kim},
  \citenamefont {Pan}, \citenamefont {Ramesh},\ and\ \citenamefont
  {Salahuddin}}]{kimapl2018}%
  \BibitemOpen
  \bibfield  {author} {\bibinfo {author} {\bibfnamefont {A.~I.}\ \bibnamefont
  {Khan}}, \bibinfo {author} {\bibfnamefont {D.}~\bibnamefont {Bhowmik}},
  \bibinfo {author} {\bibfnamefont {P.}~\bibnamefont {Yu}}, \bibinfo {author}
  {\bibfnamefont {S.~J.}\ \bibnamefont {Kim}}, \bibinfo {author} {\bibfnamefont
  {X.}~\bibnamefont {Pan}}, \bibinfo {author} {\bibfnamefont {R.}~\bibnamefont
  {Ramesh}}, \ and\ \bibinfo {author} {\bibfnamefont {S.}~\bibnamefont
  {Salahuddin}},\ }\href {\doibase 10.1063/1.5052012} {\bibfield  {journal}
  {\bibinfo  {journal} {Appl. Phys. Lett.}\ }\textbf {\bibinfo {volume}
  {113}},\ \bibinfo {pages} {182903} (\bibinfo {year} {2018})}\BibitemShut
  {NoStop}%
\bibitem [{\citenamefont {Salahuddin}\ and\ \citenamefont
  {Datta}(2008)}]{sayeefacsnano2008}%
  \BibitemOpen
  \bibfield  {author} {\bibinfo {author} {\bibfnamefont {S.}~\bibnamefont
  {Salahuddin}}\ and\ \bibinfo {author} {\bibfnamefont {S.}~\bibnamefont
  {Datta}},\ }\href {\doibase 10.1021/nl071804g} {\bibfield  {journal}
  {\bibinfo  {journal} {Nano Lett.}\ }\textbf {\bibinfo {volume} {8}},\
  \bibinfo {pages} {405} (\bibinfo {year} {2008})}\BibitemShut {NoStop}%
\bibitem [{\citenamefont {Khan}\ \emph {et~al.}(2014)\citenamefont {Khan},
  \citenamefont {Chatterjee}, \citenamefont {Wang}, \citenamefont {Drapcho},
  \citenamefont {You}, \citenamefont {Serrao}, \citenamefont {Bakaul},
  \citenamefont {Ramesh},\ and\ \citenamefont {Salahuddin}}]{asifnature2014}%
  \BibitemOpen
  \bibfield  {author} {\bibinfo {author} {\bibfnamefont {A.~I.}\ \bibnamefont
  {Khan}}, \bibinfo {author} {\bibfnamefont {K.}~\bibnamefont {Chatterjee}},
  \bibinfo {author} {\bibfnamefont {B.}~\bibnamefont {Wang}}, \bibinfo {author}
  {\bibfnamefont {S.}~\bibnamefont {Drapcho}}, \bibinfo {author} {\bibfnamefont
  {L.}~\bibnamefont {You}}, \bibinfo {author} {\bibfnamefont {C.}~\bibnamefont
  {Serrao}}, \bibinfo {author} {\bibfnamefont {S.~R.}\ \bibnamefont {Bakaul}},
  \bibinfo {author} {\bibfnamefont {R.}~\bibnamefont {Ramesh}}, \ and\ \bibinfo
  {author} {\bibfnamefont {S.}~\bibnamefont {Salahuddin}},\ }\href {\doibase
  10.1038/nmat4148} {\bibfield  {journal} {\bibinfo  {journal} {Nature
  Materials}\ }\textbf {\bibinfo {volume} {14}},\ \bibinfo {pages} {182}
  (\bibinfo {year} {2014})}\BibitemShut {NoStop}%
\bibitem [{\citenamefont {Khan}\ \emph {et~al.}(2011)\citenamefont {Khan},
  \citenamefont {Bhowmik}, \citenamefont {Yu}, \citenamefont {Kim},
  \citenamefont {Pan}, \citenamefont {Ramesh},\ and\ \citenamefont
  {Salahuddin}}]{asifapl2014}%
  \BibitemOpen
  \bibfield  {author} {\bibinfo {author} {\bibfnamefont {A.~I.}\ \bibnamefont
  {Khan}}, \bibinfo {author} {\bibfnamefont {D.}~\bibnamefont {Bhowmik}},
  \bibinfo {author} {\bibfnamefont {P.}~\bibnamefont {Yu}}, \bibinfo {author}
  {\bibfnamefont {S.~J.}\ \bibnamefont {Kim}}, \bibinfo {author} {\bibfnamefont
  {X.}~\bibnamefont {Pan}}, \bibinfo {author} {\bibfnamefont {R.}~\bibnamefont
  {Ramesh}}, \ and\ \bibinfo {author} {\bibfnamefont {S.}~\bibnamefont
  {Salahuddin}},\ }\href {\doibase 10.1063/1.3634072} {\bibfield  {journal}
  {\bibinfo  {journal} {Appl. Phys. Lett.}\ }\textbf {\bibinfo {volume} {99}},\
  \bibinfo {pages} {113501} (\bibinfo {year} {2011})}\BibitemShut {NoStop}%
\bibitem [{\citenamefont {Saha}\ \emph {et~al.}(2017)\citenamefont {Saha},
  \citenamefont {Sharma}, \citenamefont {Dabo}, \citenamefont {Datta},\ and\
  \citenamefont {Gupta}}]{sahaiedm2017}%
  \BibitemOpen
  \bibfield  {author} {\bibinfo {author} {\bibfnamefont {A.~K.}\ \bibnamefont
  {Saha}}, \bibinfo {author} {\bibfnamefont {P.}~\bibnamefont {Sharma}},
  \bibinfo {author} {\bibfnamefont {I.}~\bibnamefont {Dabo}}, \bibinfo {author}
  {\bibfnamefont {S.}~\bibnamefont {Datta}}, \ and\ \bibinfo {author}
  {\bibfnamefont {S.~K.}\ \bibnamefont {Gupta}},\ }in\ \href {\doibase
  10.1109/IEDM.2017.8268385} {\emph {\bibinfo {booktitle} {2017 IEEE
  International Electron Devices Meeting (IEDM)}}}\ (\bibinfo {year} {2017})\
  pp.\ \bibinfo {pages} {13.5.1--13.5.4}\BibitemShut {NoStop}%
\bibitem [{\citenamefont {Trentzsch}\ \emph {et~al.}(2016)\citenamefont
  {Trentzsch}, \citenamefont {Flachowsky}, \citenamefont {Richter},
  \citenamefont {Paul}, \citenamefont {Reimer}, \citenamefont {Utess},
  \citenamefont {Jansen}, \citenamefont {Mulaosmanovic}, \citenamefont
  {Müller}, \citenamefont {Slesazeck}, \citenamefont {Ocker}, \citenamefont
  {Noack}, \citenamefont {Müller}, \citenamefont {Polakowski}, \citenamefont
  {Schreiter}, \citenamefont {Beyer}, \citenamefont {Mikolajick},\ and\
  \citenamefont {Rice}}]{trentiedm2016}%
  \BibitemOpen
  \bibfield  {author} {\bibinfo {author} {\bibfnamefont {M.}~\bibnamefont
  {Trentzsch}}, \bibinfo {author} {\bibfnamefont {S.}~\bibnamefont
  {Flachowsky}}, \bibinfo {author} {\bibfnamefont {R.}~\bibnamefont {Richter}},
  \bibinfo {author} {\bibfnamefont {J.}~\bibnamefont {Paul}}, \bibinfo {author}
  {\bibfnamefont {B.}~\bibnamefont {Reimer}}, \bibinfo {author} {\bibfnamefont
  {D.}~\bibnamefont {Utess}}, \bibinfo {author} {\bibfnamefont
  {S.}~\bibnamefont {Jansen}}, \bibinfo {author} {\bibfnamefont
  {H.}~\bibnamefont {Mulaosmanovic}}, \bibinfo {author} {\bibfnamefont
  {S.}~\bibnamefont {Müller}}, \bibinfo {author} {\bibfnamefont
  {S.}~\bibnamefont {Slesazeck}}, \bibinfo {author} {\bibfnamefont
  {J.}~\bibnamefont {Ocker}}, \bibinfo {author} {\bibfnamefont
  {M.}~\bibnamefont {Noack}}, \bibinfo {author} {\bibfnamefont
  {J.}~\bibnamefont {Müller}}, \bibinfo {author} {\bibfnamefont
  {P.}~\bibnamefont {Polakowski}}, \bibinfo {author} {\bibfnamefont
  {J.}~\bibnamefont {Schreiter}}, \bibinfo {author} {\bibfnamefont
  {S.}~\bibnamefont {Beyer}}, \bibinfo {author} {\bibfnamefont
  {T.}~\bibnamefont {Mikolajick}}, \ and\ \bibinfo {author} {\bibfnamefont
  {B.}~\bibnamefont {Rice}},\ }in\ \href {\doibase 10.1109/IEDM.2016.7838397}
  {\emph {\bibinfo {booktitle} {2016 IEEE International Electron Devices
  Meeting (IEDM)}}}\ (\bibinfo {year} {2016})\ pp.\ \bibinfo {pages}
  {11.5.1--11.5.4}\BibitemShut {NoStop}%
\bibitem [{\citenamefont {Chatterjee}\ \emph {et~al.}(2017)\citenamefont
  {Chatterjee}, \citenamefont {Kim}, \citenamefont {Karbasian}, \citenamefont
  {Tan}, \citenamefont {Yadav}, \citenamefont {Khan}, \citenamefont {Hu},\ and\
  \citenamefont {Salahuddin}}]{chatedl2017}%
  \BibitemOpen
  \bibfield  {author} {\bibinfo {author} {\bibfnamefont {K.}~\bibnamefont
  {Chatterjee}}, \bibinfo {author} {\bibfnamefont {S.}~\bibnamefont {Kim}},
  \bibinfo {author} {\bibfnamefont {G.}~\bibnamefont {Karbasian}}, \bibinfo
  {author} {\bibfnamefont {A.~J.}\ \bibnamefont {Tan}}, \bibinfo {author}
  {\bibfnamefont {A.~K.}\ \bibnamefont {Yadav}}, \bibinfo {author}
  {\bibfnamefont {A.~I.}\ \bibnamefont {Khan}}, \bibinfo {author}
  {\bibfnamefont {C.}~\bibnamefont {Hu}}, \ and\ \bibinfo {author}
  {\bibfnamefont {S.}~\bibnamefont {Salahuddin}},\ }\href {\doibase
  10.1109/LED.2017.2748992} {\bibfield  {journal} {\bibinfo  {journal} {IEEE
  Electron Device Letters}\ }\textbf {\bibinfo {volume} {38}},\ \bibinfo
  {pages} {1379} (\bibinfo {year} {2017})}\BibitemShut {NoStop}%
\bibitem [{\citenamefont {Mulaosmanovic}\ \emph {et~al.}(2017)\citenamefont
  {Mulaosmanovic}, \citenamefont {Ocker}, \citenamefont {Müller},
  \citenamefont {Noack}, \citenamefont {Müller}, \citenamefont {Polakowski},
  \citenamefont {Mikolajick},\ and\ \citenamefont
  {Slesazeck}}]{mulavlsitech2017}%
  \BibitemOpen
  \bibfield  {author} {\bibinfo {author} {\bibfnamefont {H.}~\bibnamefont
  {Mulaosmanovic}}, \bibinfo {author} {\bibfnamefont {J.}~\bibnamefont
  {Ocker}}, \bibinfo {author} {\bibfnamefont {S.}~\bibnamefont {Müller}},
  \bibinfo {author} {\bibfnamefont {M.}~\bibnamefont {Noack}}, \bibinfo
  {author} {\bibfnamefont {J.}~\bibnamefont {Müller}}, \bibinfo {author}
  {\bibfnamefont {P.}~\bibnamefont {Polakowski}}, \bibinfo {author}
  {\bibfnamefont {T.}~\bibnamefont {Mikolajick}}, \ and\ \bibinfo {author}
  {\bibfnamefont {S.}~\bibnamefont {Slesazeck}},\ }in\ \href {\doibase
  10.23919/VLSIT.2017.7998165} {\emph {\bibinfo {booktitle} {2017 Symposium on
  VLSI Technology}}}\ (\bibinfo {year} {2017})\ pp.\ \bibinfo {pages}
  {T176--T177}\BibitemShut {NoStop}%
\bibitem [{\citenamefont {Jerry}\ \emph {et~al.}(2017)\citenamefont {Jerry},
  \citenamefont {Chen}, \citenamefont {Zhang}, \citenamefont {Sharma},
  \citenamefont {Ni}, \citenamefont {Yu},\ and\ \citenamefont
  {Datta}}]{jerryiedm2017}%
  \BibitemOpen
  \bibfield  {author} {\bibinfo {author} {\bibfnamefont {M.}~\bibnamefont
  {Jerry}}, \bibinfo {author} {\bibfnamefont {P.}~\bibnamefont {Chen}},
  \bibinfo {author} {\bibfnamefont {J.}~\bibnamefont {Zhang}}, \bibinfo
  {author} {\bibfnamefont {P.}~\bibnamefont {Sharma}}, \bibinfo {author}
  {\bibfnamefont {K.}~\bibnamefont {Ni}}, \bibinfo {author} {\bibfnamefont
  {S.}~\bibnamefont {Yu}}, \ and\ \bibinfo {author} {\bibfnamefont
  {S.}~\bibnamefont {Datta}},\ }in\ \href {\doibase 10.1109/IEDM.2017.8268338}
  {\emph {\bibinfo {booktitle} {2017 IEEE International Electron Devices
  Meeting (IEDM)}}}\ (\bibinfo {year} {2017})\ pp.\ \bibinfo {pages}
  {6.2.1--6.2.4}\BibitemShut {NoStop}%
\bibitem [{\citenamefont {Indiveri}\ and\ \citenamefont
  {Liu}(2015)}]{indipieee2015}%
  \BibitemOpen
  \bibfield  {author} {\bibinfo {author} {\bibfnamefont {G.}~\bibnamefont
  {Indiveri}}\ and\ \bibinfo {author} {\bibfnamefont {S.}~\bibnamefont {Liu}},\
  }\href {\doibase 10.1109/JPROC.2015.2444094} {\bibfield  {journal} {\bibinfo
  {journal} {Proceedings of the IEEE}\ }\textbf {\bibinfo {volume} {103}},\
  \bibinfo {pages} {1379} (\bibinfo {year} {2015})}\BibitemShut {NoStop}%
\bibitem [{\citenamefont {Mulaosmanovic}\ \emph
  {et~al.}(2018{\natexlab{a}})\citenamefont {Mulaosmanovic}, \citenamefont
  {Mikolajick}, ,\ and\ \citenamefont {Slesazeck}}]{mulaacsnano2018}%
  \BibitemOpen
  \bibfield  {author} {\bibinfo {author} {\bibfnamefont {H.}~\bibnamefont
  {Mulaosmanovic}}, \bibinfo {author} {\bibfnamefont {T.}~\bibnamefont
  {Mikolajick}}, , \ and\ \bibinfo {author} {\bibfnamefont {S.}~\bibnamefont
  {Slesazeck}},\ }\href {\doibase 10.1021/acsami.8b08967} {\bibfield  {journal}
  {\bibinfo  {journal} {ACS Appl. Mater. Interfaces}\ }\textbf {\bibinfo
  {volume} {10}},\ \bibinfo {pages} {1944} (\bibinfo {year}
  {2018}{\natexlab{a}})}\BibitemShut {NoStop}%
\bibitem [{\citenamefont {Ni}\ \emph {et~al.}(2018)\citenamefont {Ni},
  \citenamefont {Grisafe}, \citenamefont {Chakraborty}, \citenamefont {Saha},
  \citenamefont {Dutta}, \citenamefont {Jerry}, \citenamefont {Smith},
  \citenamefont {Gupta},\ and\ \citenamefont {Datta}}]{kaiiedm12018}%
  \BibitemOpen
  \bibfield  {author} {\bibinfo {author} {\bibfnamefont {K.}~\bibnamefont
  {Ni}}, \bibinfo {author} {\bibfnamefont {B.}~\bibnamefont {Grisafe}},
  \bibinfo {author} {\bibfnamefont {W.}~\bibnamefont {Chakraborty}}, \bibinfo
  {author} {\bibfnamefont {A.~K.}\ \bibnamefont {Saha}}, \bibinfo {author}
  {\bibfnamefont {S.}~\bibnamefont {Dutta}}, \bibinfo {author} {\bibfnamefont
  {M.}~\bibnamefont {Jerry}}, \bibinfo {author} {\bibfnamefont {J.~A.}\
  \bibnamefont {Smith}}, \bibinfo {author} {\bibfnamefont {S.}~\bibnamefont
  {Gupta}}, \ and\ \bibinfo {author} {\bibfnamefont {S.}~\bibnamefont
  {Datta}},\ }in\ \href@noop {} {\emph {\bibinfo {booktitle} {2018 IEEE
  International Electron Devices Meeting (IEDM)}}}\ (\bibinfo {year} {2018})\
  pp.\ \bibinfo {pages} {16.1.1--16.1.4}\BibitemShut {NoStop}%
\bibitem [{\citenamefont {Mulaosmanovic}\ \emph
  {et~al.}(2018{\natexlab{b}})\citenamefont {Mulaosmanovic}, \citenamefont
  {Chicca}, \citenamefont {Bertele}, \citenamefont {Mikolajickac},\ and\
  \citenamefont {Slesazecka}}]{mulananoscale2018}%
  \BibitemOpen
  \bibfield  {author} {\bibinfo {author} {\bibfnamefont {H.}~\bibnamefont
  {Mulaosmanovic}}, \bibinfo {author} {\bibfnamefont {E.}~\bibnamefont
  {Chicca}}, \bibinfo {author} {\bibfnamefont {M.}~\bibnamefont {Bertele}},
  \bibinfo {author} {\bibfnamefont {T.}~\bibnamefont {Mikolajickac}}, \ and\
  \bibinfo {author} {\bibfnamefont {S.}~\bibnamefont {Slesazecka}},\ }\href
  {\doibase 10.1039/C8NR07135G} {\bibfield  {journal} {\bibinfo  {journal}
  {Nanoscale}\ }\textbf {\bibinfo {volume} {10}},\ \bibinfo {pages} {21755}
  (\bibinfo {year} {2018}{\natexlab{b}})}\BibitemShut {NoStop}%
\bibitem [{\citenamefont {Li}\ \emph {et~al.}(2001)\citenamefont {Li},
  \citenamefont {Hu}, \citenamefont {Liu}, ,\ and\ \citenamefont
  {Chen}}]{liapl2001}%
  \BibitemOpen
  \bibfield  {author} {\bibinfo {author} {\bibfnamefont {Y.~L.}\ \bibnamefont
  {Li}}, \bibinfo {author} {\bibfnamefont {S.~Y.}\ \bibnamefont {Hu}}, \bibinfo
  {author} {\bibfnamefont {Z.~K.}\ \bibnamefont {Liu}}, , \ and\ \bibinfo
  {author} {\bibfnamefont {L.~Q.}\ \bibnamefont {Chen}},\ }\href {\doibase
  10.1063/1.1377855} {\bibfield  {journal} {\bibinfo  {journal} {Appl. Phys.
  Lett.}\ }\textbf {\bibinfo {volume} {78}},\ \bibinfo {pages} {3878} (\bibinfo
  {year} {2001})}\BibitemShut {NoStop}%
\bibitem [{\citenamefont {Rabe}, \citenamefont {Ahn},\ and\ \citenamefont
  {Triscone}(2007)}]{karin}%
  \BibitemOpen
  \bibfield  {author} {\bibinfo {author} {\bibfnamefont {K.~M.}\ \bibnamefont
  {Rabe}}, \bibinfo {author} {\bibfnamefont {C.~H.}\ \bibnamefont {Ahn}}, \
  and\ \bibinfo {author} {\bibfnamefont {J.-M.}\ \bibnamefont {Triscone}},\
  }\href {\doibase 10.1007/978-3-540-34591-6} {\emph {\bibinfo {title} {Physics
  of Ferroelectrics - A Modern Perspective}}}\ (\bibinfo  {publisher}
  {Springer},\ \bibinfo {year} {2007})\BibitemShut {NoStop}%
\bibitem [{\citenamefont {Nambu}\ and\ \citenamefont {Sagala}(1994)}]{Nambu}%
  \BibitemOpen
  \bibfield  {author} {\bibinfo {author} {\bibfnamefont {S.}~\bibnamefont
  {Nambu}}\ and\ \bibinfo {author} {\bibfnamefont {D.~A.}\ \bibnamefont
  {Sagala}},\ }\href {\doibase 10.1103/PhysRevB.50.5838} {\bibfield  {journal}
  {\bibinfo  {journal} {Phys. Rev. B}\ }\textbf {\bibinfo {volume} {50}},\
  \bibinfo {pages} {5838} (\bibinfo {year} {1994})}\BibitemShut {NoStop}%
\bibitem [{\citenamefont {Lynden-Bell}(1995)}]{bell}%
  \BibitemOpen
  \bibfield  {author} {\bibinfo {author} {\bibfnamefont {R.}~\bibnamefont
  {Lynden-Bell}},\ }\href {\doibase 10.1080/00268979500102791} {\bibfield
  {journal} {\bibinfo  {journal} {Molecular Physics}\ }\textbf {\bibinfo
  {volume} {86}},\ \bibinfo {pages} {1353} (\bibinfo {year}
  {1995})}\BibitemShut {NoStop}%
\bibitem [{\citenamefont {Elder}, \citenamefont {Gould},\ and\ \citenamefont
  {Tobochnik}(1993)}]{ken}%
  \BibitemOpen
  \bibfield  {author} {\bibinfo {author} {\bibfnamefont {K.}~\bibnamefont
  {Elder}}, \bibinfo {author} {\bibfnamefont {H.}~\bibnamefont {Gould}}, \ and\
  \bibinfo {author} {\bibfnamefont {J.}~\bibnamefont {Tobochnik}},\ }\href
  {\doibase 10.1063/1.4823138} {\bibfield  {journal} {\bibinfo  {journal}
  {Computers in Physics}\ }\textbf {\bibinfo {volume} {7}},\ \bibinfo {pages}
  {27} (\bibinfo {year} {1993})}\BibitemShut {NoStop}%
\bibitem [{\citenamefont {Bandyopadhyay}, \citenamefont {Ray},\ and\
  \citenamefont {Gopalan}(2006)}]{Band}%
  \BibitemOpen
  \bibfield  {author} {\bibinfo {author} {\bibfnamefont {A.~K.}\ \bibnamefont
  {Bandyopadhyay}}, \bibinfo {author} {\bibfnamefont {P.~C.}\ \bibnamefont
  {Ray}}, \ and\ \bibinfo {author} {\bibfnamefont {V.}~\bibnamefont
  {Gopalan}},\ }\href {http://stacks.iop.org/0953-8984/18/i=16/a=016}
  {\bibfield  {journal} {\bibinfo  {journal} {Journal of Physics: Condensed
  Matter}\ }\textbf {\bibinfo {volume} {18}},\ \bibinfo {pages} {4093}
  (\bibinfo {year} {2006})}\BibitemShut {NoStop}%
\bibitem [{\citenamefont {Cano}\ and\ \citenamefont {Jiménez}(2010)}]{cano}%
  \BibitemOpen
  \bibfield  {author} {\bibinfo {author} {\bibfnamefont {A.}~\bibnamefont
  {Cano}}\ and\ \bibinfo {author} {\bibfnamefont {D.}~\bibnamefont
  {Jiménez}},\ }\href {\doibase 10.1063/1.3494533} {\bibfield  {journal}
  {\bibinfo  {journal} {Appl. Phys. Lett.}\ }\textbf {\bibinfo {volume} {97}},\
  \bibinfo {pages} {133509} (\bibinfo {year} {2010})}\BibitemShut {NoStop}%
\bibitem [{\citenamefont {Emelyanov}\ \emph {et~al.}(2002)\citenamefont
  {Emelyanov}, \citenamefont {Pertsev}, \citenamefont {Hoffmann-Eifert},
  \citenamefont {B{\"o}ttger},\ and\ \citenamefont {Waser}}]{Emelyanov2002}%
  \BibitemOpen
  \bibfield  {author} {\bibinfo {author} {\bibfnamefont {A.}~\bibnamefont
  {Emelyanov}}, \bibinfo {author} {\bibfnamefont {N.}~\bibnamefont {Pertsev}},
  \bibinfo {author} {\bibfnamefont {S.}~\bibnamefont {Hoffmann-Eifert}},
  \bibinfo {author} {\bibfnamefont {U.}~\bibnamefont {B{\"o}ttger}}, \ and\
  \bibinfo {author} {\bibfnamefont {R.}~\bibnamefont {Waser}},\ }\href
  {\doibase 10.1023/A:1021665300233} {\bibfield  {journal} {\bibinfo  {journal}
  {Journal of Electroceramics}\ }\textbf {\bibinfo {volume} {9}},\ \bibinfo
  {pages} {5} (\bibinfo {year} {2002})}\BibitemShut {NoStop}%
\bibitem [{\citenamefont {Lee}\ \emph {et~al.}(1999)\citenamefont {Lee},
  \citenamefont {Park}, \citenamefont {Park}, \citenamefont {Lee},\ and\
  \citenamefont {Joo}}]{joo_1999}%
  \BibitemOpen
  \bibfield  {author} {\bibinfo {author} {\bibfnamefont {J.-S.}\ \bibnamefont
  {Lee}}, \bibinfo {author} {\bibfnamefont {E.-C.}\ \bibnamefont {Park}},
  \bibinfo {author} {\bibfnamefont {J.-H.}\ \bibnamefont {Park}}, \bibinfo
  {author} {\bibfnamefont {B.-I.}\ \bibnamefont {Lee}}, \ and\ \bibinfo
  {author} {\bibfnamefont {S.-K.}\ \bibnamefont {Joo}},\ }\href {\doibase
  10.1557/PROC-596-315} {\bibfield  {journal} {\bibinfo  {journal} {MRS
  Proceedings}\ }\textbf {\bibinfo {volume} {596}},\ \bibinfo {pages} {315}
  (\bibinfo {year} {1999})}\BibitemShut {NoStop}%
\end{thebibliography}%

\end{document}